\documentclass[12pt]{article}
\usepackage[colorlinks=false, urlcolor=citecolor, linkcolor=citecolor, citecolor=citecolor]{hyperref}

\usepackage{textcomp}
\usepackage{amsfonts}
\usepackage{amssymb}
\usepackage{verbatim}

\usepackage{mathrsfs}
\usepackage{multirow}

\usepackage{epsfig}
\usepackage{amsmath}
\usepackage{mathtools}
\usepackage{multirow}
\usepackage{amsthm}
\usepackage{lscape}
\usepackage{natbib}
\usepackage{graphicx}

\usepackage{hhline}

\newtheorem{theorem}{Theorem}
\newtheorem{corollary}[theorem]{Corollary}

\begin{document}

\title{Combining parametric, semi-parametric, and non-parametric survival models with stacked survival models}

\author{ANDREW WEY$^{1}$, JOHN CONNETT, KYLE RUDSER}

\date{ }


\maketitle


\begin{abstract}
{For estimating conditional survival functions, non-parametric estimators can be preferred to parametric and semi-parametric estimators due to relaxed assumptions that enable robust estimation.  Yet, even when misspecified, parametric and semi-parametric estimators can possess better operating characteristics in small sample sizes due to smaller variance than non-parametric estimators.  Fundamentally, this is a bias-variance tradeoff situation in that the sample size is not large enough to take advantage of the low bias of non-parametric estimation.  Stacked survival models estimate an optimally weighted combination of models that can span parametric, semi-parametric, and non-parametric models by minimizing prediction error.  An extensive simulation study demonstrates that stacked survival models consistently perform well across a wide range of scenarios by adaptively balancing the strengths and weaknesses of individual candidate survival models.  In addition, stacked survival models perform as good as, or better than, the model selected through cross-validation.  Lastly, stacked survival models are applied to a well-known German breast cancer study.}

\end{abstract}

\footnote{University of Hawaii and University of Minnesota}

\section{Introduction}
Survival function estimation has long been a major component of survival analysis.  Yet estimation of conditional survival functions, i.e., survival functions that depend on covariate values, remains a challenging problem.  A common semi-parametric approach combines the Cox proportional hazard model with a baseline hazard estimate, e.g., see \citet{kalbfleisch:prentice:2002}.  However, if the functional form is misspecified or the proportional hazards assumption is violated, then this approach may perform poorly.  In terms of the bias-variance tradeoff, the Cox model, and other parametric models, achieve low variance by making distributional and functional form assumptions.  If the assumptions are approximately correct, then the bias term is small and the parametric and semi-parametric models perform well.  On the other hand, if the assumptions are badly violated, then the bias term can be large and the models perform poorly. 

Many non-parametric methods have been proposed to overcome the bias induced by violated assumptions.  For example, \citet{kooperberg:stone:truong:1995} proposes a flexible spline approach for the log-hazard that encompasses more than a proportional hazards model.  Alternatively, tree-based approaches have been considered by several authors \citep{ishwaran:others:2008, bouhamad:others:2011, zhu:kosorok:2012}.  Despite possessing low bias in a wide variety of situations, non-parametric estimators suffer from high variance and can require a large sample size to perform well.  This can lead to surprising situations where misspecified parametric models perform better than non-parametric estimators.  Specifically, the effect of bias of misspecified parametric models is smaller than the  effect of variance of non-parametric estimators, i.e., the bias-variance trade-off.

This article pursues a flexible estimator of a conditional survival function, i.e., an estimator that performs well when parametric assumptions are approximately correct while also maintaining robustness when parametric assumptions are violated.  Traditionally, a single conditional survival function estimator is chosen from a set of candidate models, e.g., using an information criterion \citep{kooperberg:stone:truong:1995} or through cross-validation.  Rather than select a single survival model, our goal is to estimate an optimally weighted combination of several survival models.

A variety of approaches that combine several models, often referred to as ensembles, have been explored in the uncensored setting.  One approach, called ``stacking,'' determines the optimally weighted average of several models by minimizing predicted error.  \citet{wolpert:1992} introduced stacking in the context of neural networks, while \citet{breiman:stacking:1996} extended the idea to uncensored regression models and showed that stacking could improve prediction error. In particular, \citet{breiman:stacking:1996} found that combining fundamentally different regression models, e.g., ridge regression and subset regression, had the largest reduction in prediction error.  \citet{leblanc:tibshirani:1996} found stacking with a constraint of non-negative weights to be an efficient way to combine models.  \citet{vanderLaan:polley:hubbard:2007} independently developed uncensored stacking as the `Super Learner' algorithm, and presented results regarding the stacked estimator's rate of convergence.  More recently, \citet{boonstra:taylor:mukherjee:2013} used stacking to improve prediction when incorporating different generation sequencing information in high dimensional genome analysis.  

Stacking models in a censored data setting presents additional challenges.  \citet{polley:vanderLaan:2011} mention stacking within a general censored data framework and provide an example for hazard function estimation with discrete survival times.  This paper differs in two notable ways.  First, we focus on estimating conditional survival functions with continuous survival times rather than a hazard function.  This requires a different loss function that is tailored directly to estimating survival functions.  We are particularly interested in the conditional survival function due to its role in many survival analysis methods; see the last paragraph of Section~7 for several examples.  We also pursue the potential advantages of stacking parametric, semi-parametric, and non-parametric estimators.  In particular, we show that stacked survival models perform well by giving weight to approximately correct parametric models, while shifting weight to non-parametric estimators when assumptions are violated.  This allows stacked survival models to outperform the single model selected via cross-validation and, in some situations, outperform every individual model considered in the stacking procedure.  We believe that combining parametric, semi-parametric, and non-parametric estimators is the biggest advantage of stacked survival models.  

The remainder of the manuscript is organized as follows: stacked survival models are proposed in Section~2.  Section~\ref{bias:var:sec} investigates the mean-squared error of stacked survival models with some asymptotic properties presented in Section~\ref{asymp:sec}.  Section~\ref{sims:sec} investigates the finite sample performance through an extensive simulation study.  Stacked survival models are then applied to the German breast cancer study data set in Section~6, with concluding remarks presented in Section~7.

\section{Stacking Survival Models}
Throughout the paper, random variables and observed variables are distinguished by capital and lower case letters, respectively.  Our objective is to estimate the survival function of the event time random variable $T$ that depends on $p$ baseline covariates $\boldsymbol{x}$, i.e., $S_o(t|\boldsymbol{x}) = P(T > t | \boldsymbol{x})$.  In survival analysis, $T$ may be only partially observed due to a censoring random variable $C$ that may also depend on $\boldsymbol{x}$.  Define the conditional survival function of the censoring distribution as $G(t|\boldsymbol{x}) = P(C > t | \boldsymbol{x})$.  We assume throughout that the event time and censoring random variables are conditionally independent, i.e., $T \bot C | \boldsymbol{x}$.  The observed time is $y_i = \text{min}(t_{i},c_{i})$, and $\delta_{i} = I(t_{i} < c_{i})$ indicates whether an event was observed.  Hence a sample of right censored survival data of size $n$ consists of triplets $\{y_{i}, \delta_{i}, \boldsymbol{x}_{i} \}$, $i = 1,...,n$.  Using the observed triplets, we can construct, for example, an estimate of the event time survival function from each of $m$ candidate models with the $k^{th}$ estimate denoted as $\hat{S}_k(t|\boldsymbol{x})$.  

In order to combine several predictors, we need a loss function that is tailored to survival functions.  Our approach uses the Brier Score [$BS(t)$], which measures the squared error of a survival function at a given time point.  In the absence of censoring, $BS(t)$ is defined as
\begin{eqnarray}
\label{full:brier:score:t}
BS(t) &=&  \frac{1}{n} \sum_{i = 1}^{n} \{ Z_i(t) - \hat{S}(t | \boldsymbol{x}_i) \}^2,
\end{eqnarray}
where $Z_i(t) = I(t_i > t)$.  Note that $t$ denotes a chosen time point, while $t_i$ (with the subscript) denotes the event time for the $i^{th}$ observation and may not be observed due to censoring.  

Unfortunately, right-censoring implies that equation (\ref{full:brier:score:t}) is only partially observed; that is, $Z_i(t)$ is undefined for censored observations when $y_i > t$.  To correct this issue, we use inverse probability-of-censoring weights (IPCW) to account for the probability of an observation being censored \citep{lostritto:strawderman:molinaro:2012}.  In particular, the `inverse probability-of-censoring weighted Brier Score' at time $t$ [$IPCW\text{-}BS(t)$] can be written as
\begin{eqnarray}
\label{brier:score:t}
IPCW\text{-}BS(t) &=&  \frac{1}{n} \sum_{i = 1}^{n} \frac{\Delta_i(t)}{G(T_i(t) | \boldsymbol{x}_i)} \times \{ Z_i(t) - \hat{S}(t | \boldsymbol{x}_i) \}^2,
\end{eqnarray}
where $T_i(t) = \min\{t_i, t\}$, $\Delta_i(t) = I(\min \{t_i, t\} < c_i)$, and $G(\cdot | \boldsymbol{x}_i)$ is the conditional survival function of the censoring distribution, which is estimated by a marginal Kaplan-Meier throughout the rest of the paper.  From a technical point of view, the inverse probability-of-censoring weights ensure that the expectations of the equations (\ref{full:brier:score:t}) and (\ref{brier:score:t}) are the same (assuming that the estimator of the censoring distribution is uniformly consistent).  Thus the true conditional survival function, $S_o(t|\boldsymbol{x})$, is the minimizer of $E\{ IPCW\text{-}BS(t) \}$ (see the Supplementary Materials).

There are several points that are helpful to note for understanding the calculation of the $IPCW\text{-}BS(t)$.   The values of $T_i(t)$ and $\Delta_i(t)$ depend on $t$ and the censoring status of the $i^{th}$ observation.  For each uncensored observation, the value of $G(T_i(t) | \boldsymbol{x}_i)$, and therefore the calculation of the weight, depends on whether the event has occurred by time $t$.  For censored observations, there are two possible situations at a given time point $t$:
\begin{itemize}
	\item If $c_i > t$, then $T_i(t) = t$ and $\Delta_i(t) = I(t < c_i) = 1$.
	\item If $c_i < t$, then $\Delta_i(t) = 0$ (since $c_i < t_i$ for censored observations) and hence $\frac{\Delta_i(t)}{G(T_i(t) | \boldsymbol{x}_i)} =~0$.
\end{itemize}
Thus, for a fixed time $t$, censored observations with $c_i > t$ will contribute to the Brier Score, while censored observations with $c_i < t$ will still contribute to the Brier Score but only indirectly through the estimation of the censoring distribution.

Since the goal is to estimate the entire conditional survival function, the Brier Score is minimized over a set of time points, say $t_1,...,t_s$.  This implies the following weighted least squares problem with the additional constraints that $\sum_{k=1}^m \hat{\alpha}_k = 1$, which is required for the theoretical results, and $\hat{\alpha}_k \geq 0$ for all $k = 1,...,m$, which has been shown to improve performance in the uncensored setting \citep{breiman:stacking:1996, leblanc:tibshirani:1996},
\begin{eqnarray}
\label{bs:stacking:t}
\hat{\boldsymbol{\alpha}} &=& \text{arg} \min_{\boldsymbol{\alpha}, \alpha_k \ge 0} \sum_{r = 1}^{s} \sum_{i = 1}^{n} \frac{\Delta_i(t_r)}{\hat{G}(T_i(t_r) | \boldsymbol{x}_i)} \times \{ Z_i(t_r) - \sum_{k = 1}^{m} \alpha_k \hat{S}_k^{(-i)}(t_r | \boldsymbol{x}_i) \}^2,
\end{eqnarray}
where $\hat{S}_k^{(-i)}(t | \boldsymbol{x}_i)$ is the survival estimate from the $k^{th}$ model while leaving the $i^{th}$ observation out during the fitting process.  This ensures that stacking does not reward model complexity (i.e., does not overfit the data).  To reduce computational demands, we use five-fold cross-validation rather than n-fold cross-validation to obtain $\hat{S}_k^{(-i)}(t | \boldsymbol{x}_i)$.  In particular, the data is randomly split into five roughly equally sized sets and $\hat{S}_k^{(-i)}(t | \boldsymbol{x}_i)$ is obtained for observations in a given set by fitting the candidate survival models to the observations in the other four sets.  As such, five survival models, rather than $n$ survival models, are fit for each of the $m$ candidate survival models.  

Finally, the stacked estimate of the conditional survival function with time-independent weights is
\begin{eqnarray}
\label{stacked:surv:est}
\hat{S}(t|\boldsymbol{x}) &=& \sum_{k = 1}^{m} \hat{\alpha}_k \hat{S}_k(t | \boldsymbol{x}),
\end{eqnarray}
where $\hat{S}_k(t | \boldsymbol{x})$ is the $k^{th}$ survival model estimated with all the data.

\noindent
{\bf Remark 1.}  The Brier Score measures agreement at only one particular time.  As such, the value(s) of $t$ over which it is evaluated, i.e., $t_1,..., t_s$, have implications for performance.  In particular, care should be taken to avoid picking only very small, or very large $t$ values, though one could also consider unequal weighting or restricting to certain areas of support.  We find that nine evenly spaced quantiles of the observed event distribution works well (see the Supplementary Materials).

\noindent
{\bf Remark 2.} Time-dependent stacking, i.e., allowing the weighted combination of models to depend on time, was also considered (see the Supplementary Materials).  Though potentially adding flexibility, a major flaw of time-dependent stacking is that the conditional survival function may, at times, increase, which violates the non-increasing property of survival functions.  As such, this paper focuses on time-independent stacking.

\section{Mean-Squared Error Decomposition}
\label{bias:var:sec}
We analyze the mean-squared error of the stacked survival model.  We start by defining the mean-squared error for the stacked estimator as $\text{MSE}_{\tau}\{ \hat{S}(\cdot | \boldsymbol{x}) \} =  E \int_0^{\tau} [ \hat{S}(t | \boldsymbol{x}) - S_o(t| \boldsymbol{x})]^2dt$, where the expectation is over the random variable for the covariate space and the sampling distribution of the stacked estimator.  This definition of mean-squared error is motivated, in part, by the Brier Score.  In particular, the Supplementary Materials show that $E \int_0^{\tau} IPCW\text{-}BS(t)dt = \sigma^2 + \text{MSE}_{\tau}\{ \hat{S}(\cdot | \boldsymbol{x}) \}$, where $\sigma^2$ is irreducible prediction error.  Similar to the analysis of \citet{fumera:roli:2005}, we show in the Supplementary Materials that the mean-squared error decomposes into
\begin{eqnarray}
\label{mse:decomp}
\text{MSE}_{\tau}\{ \hat{S}(\cdot | \boldsymbol{x})\} &=& \sum_{k=1}^m \alpha_k^2 \text{MSE}_{\tau} \{ \hat{S}_k(\cdot | \boldsymbol{x})\} + E \sum_{k=1}^m \sum_{l \neq k} \alpha_k \alpha_l \int_0^{\tau} \left[ \text{Bias}\{ \hat{S}_k(t | \boldsymbol{x})\} \times \text{Bias}\{ \hat{S}_l(t | \boldsymbol{x})\} + \right. \nonumber \\
&& ~~~~~ \left.  \text{Corr}\{ \hat{S}_k(t | \boldsymbol{x}), \hat{S}_l(t | \boldsymbol{x}) \} \times \text{Var}\{ \hat{S}_k(t | \boldsymbol{x}) \}^{\frac{1}{2}} \times \text{Var}\{ \hat{S}_l(t | \boldsymbol{x}) \}^{\frac{1}{2}} \right]dt , \nonumber
\end{eqnarray}
where $\text{MSE}\{ \hat{S}_k(\cdot | \boldsymbol{x})\}$, $\text{Bias}\{ \hat{S}_k(t | \boldsymbol{x})\} = E[ \hat{S}_k(t | \boldsymbol{x}) - S_o(t| \boldsymbol{x})]$, and $\text{Var}\{ \hat{S}_k(t | \boldsymbol{x})\} = E\{ [ \hat{S}_k(t | \boldsymbol{x}) - E \hat{S}_k(t | \boldsymbol{x})]^2 \}$ are, respectively, the mean-squared error, bias at time $t$, and variance at time $t$ for the $k^{th}$ survival model in the stacking procedure, while $\text{Corr}\{ \hat{S}_k(t | \boldsymbol{x}), \hat{S}_l(t | \boldsymbol{x}) \}$ is the correlation at time $t$ between the $k^{th}$ and $l^{th}$ survival model.

The mean-squared error of the stacked estimator decomposes into two parts: a weighted combination of the mean-squared error of candidate survival models and the interaction between candidate survival models in terms of bias and correlation.  The decomposition makes it easy to show, given a set of candidate survival models, that there exists a set of stacking weights such that the stacked estimator possess as good, or better, mean-squared error as the best performing model in the set of candidate survival models.  However, this property is {\it not} guaranteed after estimating the stacking weights.  Thus, careful selection of candidate survival models is warranted.

The MSE decomposition provides insight into how features of candidate survival models impact performance of the stacked estimator.  As stated in the introduction, the motivation for stacked survival models is to obtain robustness across a wide variety of scenarios by including models from different classes, i.e., parametric, semi-parametric and non-parametric models, and with different assumptions (e.g., proportional hazards or accelerated failure time).  In this fashion, the stacked estimator may assign more weight to one model in the stack for one scenario and shift to another model, e.g., one based on different assumptions, for a different scenario.  This motivates including a set of models that ``represent'' a variety of classes and types of models, i.e., ensuring a diverse set of candidate survival models \citep{breiman:stacking:1996}.  This is also supported by the MSE decomposition as the correlation between diverse models will tend to be lower due to different assumptions.  

The next consideration is the number of models of a given type to include in the stack, e.g., the number of Cox proportional hazards models to include.  Due to numerous options regarding potential covariates and the functional form of those covariates, e.g., linear terms versus quadratic terms, there are many different Cox models that could be included.  However, models with the same distributional assumptions and similar sets of covariates are expected to have similar MSE with a rather high between-model correlation.  Since there is no guarantee that only one model among a set of highly correlated models will receive non-zero weight, the MSE decomposition suggests that the stack will perform better by excluding models with small differences in the set of considered covariates.  Further discussion, illustrative examples, and simulations are included in the Supplementary Materials.


\section{Asymptotic Properties}
\label{asymp:sec}
We show model selection and uniform consistency for the stacked estimate of the conditional survival function.  The former refers to the idea that if the set of stacked models contains uniformly consistent models, then all weight is asymptotically given to those models in the stack.  Consistent model selection implies uniform consistency as long as there is at least one uniformly consistent estimator of the conditional survival function.  Our main assumption is that there exists no weighted average of misspecified models that approaches the true survival function for every time point included in equation (\ref{bs:stacking:t}).  The Supplementary Materials contain all of the assumptions and proofs.

Let $\Omega = (0, \tau)$ be the support of interest for estimating the conditional survival function, and consider $m$ estimators for the stacking procedure.  Then
\begin{theorem}
\label{model:selection:sl}
Let $\hat{\boldsymbol{\alpha}}$ be estimated by equation (\ref{bs:stacking:t}).  Assume that models $1,...,l$, where $l < m$, are the only uniformly consistent estimators and conditions (A1)-(A3) in the Supplementary Materials hold, then $\sum_{k = 1}^l \hat{\alpha}_k \rightarrow 1$, in probability, as $n \rightarrow \infty$.
\end{theorem}
\noindent
This ensures that uniformly consistent model(s) will asymptotically receive all of the weight for the stacked conditional survival function estimate in equation (\ref{stacked:surv:est}).  There can be more than one uniformly consistent estimator, e.g., a correctly specified Weibull model and Cox model.  In the special case, when only one model is uniformly consistent, we obtain the corollary:
\begin{corollary}
\label{model:selection:s1}
If $\hat{S}_1(t|\boldsymbol{x})$ is the only uniformly consistent estimator, then $\hat{\alpha}_1\rightarrow 1$, in probability, as $n \rightarrow \infty$.
\end{corollary}
\noindent
The result of Theorem \ref{model:selection:sl} and Corollary \ref{model:selection:s1} is required for uniform consistency of the stacked estimator with time-independent weights.
\begin{theorem}
\label{uniform:consistency}
Let the stacked estimate of the conditional survival function be defined as $\hat{S}(t | \boldsymbol{x})$ in equation (\ref{stacked:surv:est}).  Assume that conditions (A1)-(A3) in the Supplementary Materials hold then, as $n \rightarrow \infty$,
\begin{eqnarray}
\sup_{t \epsilon \Omega} \sup_{\boldsymbol{x}} | \sum_{k = 1}^m \hat{\alpha}_k \hat{S}_k(t | \boldsymbol{x}) - S_o(t | \boldsymbol{x}) | \rightarrow 0. \nonumber
\end{eqnarray}
\end{theorem}
The rate of convergence of the stacked estimator is not addressed here.  However, \citet{vanderLaan:polley:hubbard:2007} showed that, in the uncensored case, the stacked estimator's risk converged at either the best rate of a correctly specified model, or slightly slower than the parametric rate.    These results are not directly applicable since the Brier Score does not measure the risk of the entire conditional survival function.  In addition, distributional results for the conditional survival function are complicated by the constrained estimation of $\boldsymbol{\alpha}$ [see Supplementary Materials for in-depth discussion].

\section{Simulations}
\label{sims:sec}
An extensive simulation study examines the finite sample performance of stacked survival models.  In particular, two settings are investigated: a moderate number of covariates (Section \ref{sims:sec:low:cen}) and a large number of covariates (Section \ref{sims:sec:high:cov}).  

The simulations are comprised of combinations of an event distribution ($d = 1, 2, 3$) and linear form of covariates ($q = 1, 2$).  The covariate distributions are multivariate normal: $\boldsymbol{x}_p \sim MVN(\boldsymbol{0}, \Sigma)$, where $\Sigma$ is the correlation matrix and for all $i, j = 1,...,p$, $\Sigma_{i,j} = \rho^{|i - j|}$ with $\rho = 0.4$ ($p$ is the vector dimension).  Section \ref{sims:sec:low:cen} has an eight-dimensional covariate space (i.e., $p = 8$), while Section \ref{sims:sec:high:cov} has a $p = 80$ dimensional covariate space.  For Section \ref{sims:sec:low:cen}, the covariate effects are $\boldsymbol{\beta} = (1, 0, -1, 0, 0.5, 0, -0.5, 0)$, while for Section \ref{sims:sec:high:cov} the first twelve covariate effects are $(1, 0, -1, 0, 0.5, 0, -0.5, 0, 0.25, 0, -0.25, 0)$ with the other $68$ effects set to zero.  Two different linear combinations are considered: $\boldsymbol{\gamma}^1 = \boldsymbol{x}_p$ and $\boldsymbol{\gamma}^2 = \Phi(4 \times \boldsymbol{x}_p)$ which imply linear and non-linear covariate effects, respectively.  The event distributions are defined as
\begin{enumerate}
	\item $T_1^{(q)} \sim \exp \{ \text{Normal}(\boldsymbol{\beta} \boldsymbol{\gamma}^q, \frac{1}{4}) \}$
	\item $T_2^{(q)} \sim \text{Weibull}(\text{scale} = \exp \{ \boldsymbol{\beta} \boldsymbol{\gamma}^q \}, \text{shape} = 1.1)$
	\item $T_3^{(q)} \sim \text{Gamma}(\text{scale} = \frac{1}{4} \exp \{ \boldsymbol{\beta} \boldsymbol{\gamma}^q \}, \text{shape} = 5)$
\end{enumerate}
Each subsection investigates every combination of the event distribution ($d$) and linear form ($q$), i.e., there are six scenarios for both Sections~5.1 and 5.2.

We compare the performance of survival models based on an approximation to the mean squared error presented in Section~\ref{bias:var:sec}, which we call integrated squared survival error (ISSE):
\begin{eqnarray}
\text{ISSE} & \approx & \frac{1}{n} \sum_{i = 1}^{n} \sum_{j = 1}^{19} (\hat{S}(t_{j} |\boldsymbol{x}_i) - S_o(t_{j} |\boldsymbol{x}_i))^2, \nonumber
\end{eqnarray}
where $t_{j}$ are a fixed set of 19 equally spaced quantiles of the survival time distribution given that an event occurs, and $S_o( \cdot |\boldsymbol{x}_i)$ is the true conditional survival function.

One comparison of interest for the stacked estimator is the model chosen through cross-validation.  We use the integrated Brier Score (IBS) as the measure of the predicted error for selecting the individual model.  In particular, the IBS for the $k^{th}$ model is defined as $\hat{\text{IBS}}_k = \int_0^{\tau} \hat{BS}_k(t)dt$, where $\tau$ is the maximum observed time and $\hat{BS}_k(t)$ is the estimated Brier Score at time $t$ for the $k^{th}$ model (with an out-of-bag estimate of the conditional survival function).  The cross-validated estimator is then defined as $\hat{S}_l(\cdot | \boldsymbol{x})$, where $l = \text{arg} \min_k \hat{\text{IBS}}_k$.

All simulations were run in R version 3.0.0 \citep{R}.  The constrained minimization problem was solved using the \texttt{alabama} package \citep{alabama:2013}.  The stacking weights, i.e., equation (\ref{bs:stacking:t}), were estimated by minimizing the Brier Score over the $0.1, 0.2, ..., 0.9$ quantiles of the observed event distribution.

\subsection{Modest Dimensional Covariate Space}
\label{sims:sec:low:cen}
This setting has relatively few covariates ($p = 8$) with a modest censoring rate ($25\%$) and sample size ($n = 200$).  This illustrates stacked survival models in a relatively straightforward scenario.  

The stacked survival models include a Weibull model and log-Normal model as parametric models, a Cox proportional hazards model with an Efron estimate of the baseline cumulative hazard function as a semi-parametric model, and random survival forests (RSF) as a non-parametric model.  The parametric and semi-parametric models only include first-order main effects and no interactions.  All of the parametric and semi-parametric models are estimated using the \texttt{survival} package in R \citep{survival}, and all of the parametric and semi-parametric models use five-fold cross-validation to estimate $\hat{S}_k^{(-i)}(t | \boldsymbol{x}_i)$.  RSF is estimated with the \texttt{randomSurvivalForest} package in R \citep{randomSurvivalForest}.  The RSF is an ensemble of $250$ trees grown using package defaults.  For RSF, $\hat{S}_k^{(-i)}(t | \boldsymbol{x}_i)$ is estimated with the out-of-bag ensemble from the \texttt{rsf} function.  The censoring distribution is a uniform distribution for all $T_d^{(q)}$: $C_{d,q} \sim  \text{Unif}(0, c(d, q))$, where $c(d, q)$ is a constant that depends on $(d, q)$ and ensures approximately $25\%$ censoring.


The log-Normal and Weibull scenarios with linear covariate effects illustrate performance when there is a correctly specified parametric or semi-parametric model in the stack.  Stacking is not expected to perform better than a correctly specified parametric model, but should still perform relatively well in such situations.  The Gamma scenario with linear covariate effects illustrates performance when there are approximately correct parametric models in the stack (e.g., a correct mean function).  The scenarios with non-linear covariate effects were designed to have badly misspecified parametric and semi-parametric models.  Due to the lack of a correctly specified parametric model, stacked survival models should perform relatively well by, in particular, assigning more weight to the non-parametric estimator: random survival forests (RSF).

Table \ref{tab:sec:low:cen} presents the results in terms of integrated squared survival error (ISSE).  Since the goal is an estimator that performs well in a wide variety of situations, the top two estimators are bolded for each scenario.  The stacked survival model, i.e., ``Stacking'', is a top two estimator for all six scenarios.  For the scenarios with non-linear covariate effects, the stacked estimator reduces the ISSE by approximately $8\%-15\%$ compared to the best single model.  In addition, the stacking procedure outperforms selecting a single model via cross-validation in every situation.

As an illustration, Table \ref{tab:low:cen:ave:wghts} presents the average stacking weights for the individual models.  For the linear scenarios, the stacking procedure gives a majority of weight to correctly specified parametric models.   The weights are more interesting for the scenarios with non-linear covariate effects.  In particular, the parametric models always receive over $40\%$ of the weight despite, at times, having $10\%$ higher ISSE than RSF.  This is a good example of stacked survival models combining misspecified parametric models and an inefficient non-parametric model to obtain an estimator that outperforms every single model considered in the stacking procedure.

\noindent
{\bf Remark 3.} Random survival forests (RSF) possess tuning parameters that influence performance, e.g., the minimum number of events in a node.  While the performance of RSF could be improved by adaptively selecting tuning parameters (e.g., by cross-validation), stacked survival models are likely to also inherit any improvement in RSF since it is included in the stack.

\subsection{Large Covariate Space}
\label{sims:sec:high:cov}
This setting has a large number of covariates ($p = 80$) relative to the sample size ($n = 200$).  The censoring distributions are the same as Section \ref{sims:sec:low:cen}.  In general, the parametric and semi-parametric models used (and stacked) in Section \ref{sims:sec:low:cen} will not perform well in large covariate spaces without regularization.  As such, they are not included for these scenarios.  We instead stack a Cox model with an $l_1$ penalty (i.e., lasso), a boosted version of the Cox model, and random survival forests (RSF).  The $l_1$ penalized version of the Cox model is fit using the R package \texttt{penalized} with the penalty parameter chosen via cross-validation \citep{penalized}.  The boosted Cox model is fit using the package \texttt{CoxBoost} in R with default tuning parameters \citep{CoxBoost}.  RSF is fit in the same manner as Section \ref{sims:sec:low:cen}.

The stacked survival model is again a top two estimator in every scenario (see Table \ref{tab:sec:high:cov}).  Relative to Section \ref{sims:sec:low:cen}, stacked survival models offer smaller improvements (e.g., approximately $1\% - 5\%$ lower ISSE compared to the cross-validated estimator).  However, the improvements in ISSE remain consistent across the scenarios.  In addition, the stacking procedure still performs as good, or better, in every scenario as the model selected via cross-validation.

\noindent
{\bf Remark 4.}  The Supplementary Materials present numerous extensions to the simulation study.  In particular, we investigate scenarios with a larger sample size, a high censoring rate, non-monotonic covariate effects, and a misspecified censoring model.  In addition, we comment on the required computational time and the influence of the out-of-bag estimator for RSF.

\section{German Breast Cancer Study}
Stacked survival models are illustrated on a well-known survival benchmark data set: the German breast cancer study (GBCS) described by \citet{hosmer:lemeshow:may:2008}, and accessible at the University of Massachusetts website for statistical software information.  There are eight covariates included in the analysis: age at diagnosis, tumor size, tumor grade, number of nodes, menopausal status, the number of progesterone receptors, the number of estrogen receptors, and hormone therapy status.  The outcome of interest is the time till death, and there is complete data on $686$ patients with approximately $75\%$ censoring.  The stacking procedure uses the same models as Section \ref{sims:sec:low:cen}.  That is, the Weibull and log-Normal model are the parametric models, the Cox proportional hazards model is the semi-parametric model, and a random survival forest is the non-parametric model.  The minimum number of deaths (for RSF) is set at $12$, which was selected by minimizing predicted error among five potential values: $3, 6, 12, 24, 48$.

We are particularly interested in the association of tumor size and the number of nodes with five-year survival.  In order to evaluate the association, the stacked survival model and each model included in the stacking procedure predicts the five-year survival rate for each patient in the study.   After predicting five-year survival, a generalized additive model with penalized B-splines for the continuous covariates [i.e., the \texttt{gam} function from the \texttt{mgcv} package \citep{wood:2006}] estimates the association of tumor size and the number of nodes with five-year survival while adjusting for the other covariates.  

\begin{figure}[hptb]
\centering
\caption{The association of tumor size (mm) and the number of nodes with five-year survival for the GBCS data set with the other covariates to their median value.  The tick marks at the bottom of the plots indicate the skewness of both covariates.}
\includegraphics[height=8cm,width=16cm]{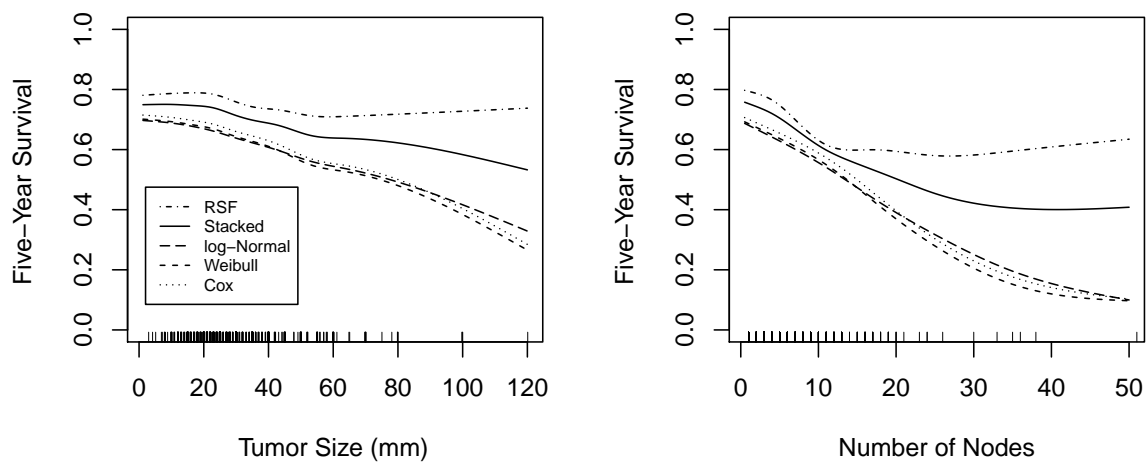}
\label{fig:gbcs:pred:surv}
\end{figure}

Figure \ref{fig:gbcs:pred:surv} presents the estimated five-year survival as a function of tumor size and the number of nodes at the median of the other covariates.  The parametric/semi-parametric models suggest worse five-year survival with increasing tumor size and number of nodes.  In contrast, RSF suggests that five-year survival dips slightly around $40mm$ for tumor size, while five-year survival for the number of nodes has a sharp early decrease but plateaus after about $10$ nodes.  The stacked survival model - which gives weight to the Weibull model ($0.06$), the Cox model ($0.38$), and RSF ($0.56$) - is a compromise between the parametric/semi-parametric models and RSF. 

The GBCS data set has a marginal five-year survival rate of $70\%$ due, in part, to a censoring rate of $75\%$.  As such, predicted five-year survival rates less than $20\%$ are surprising (i.e., the parametric/semi-parametric models for the number of nodes).  Due to the sparsity of patients with more than $20$ nodes, the low model based predicted probabilities are likely due to parametric/semi-parametric models being heavily influenced by a strong negative association with survival for patients with less than $20$ nodes ($98\%$ of patients have less than $20$ nodes) through the first order linear effect (note that the patient with over $50$ nodes was censored after two years).  In contrast, RSF does not require any linearity assumptions and is more influenced by local observations in predicting five-year survival \citep{ishwaran:others:2008}.  From this perspective, the stacked survival model is balancing model based predictions that require assumptions of linearity with locally based predictions.

\noindent
{\bf Remark 5.}  In this example, the model weights provide insight into how candidate survival models were combined to form the stacked estimate of the conditional survival function.  However, we caution against interpreting model weights as an indication of a ``correct model.''  As noted by a referee, this is particularly dangerous when two models possess similar survival functions due to potential instability in the minimization procedure.

\section{Conclusion and Future Directions}
We propose stacking survival models to flexibly estimate conditional survival functions.  Stacked survival models can combine several models, spanning the full range of parametric, semi-parametric, and non-parametric estimators.  This allows stacking to exploit the low variance of approximately correct parametric models, while maintaining the robustness of non-parametric estimators.  As illustrated in the simulation study, stacked survival models give more weight to parametric and semi-parametric models when assumptions are approximately correct, but shift weight to non-parametric estimators when assumptions are badly violated.  In this manner, stacked survival models perform well across a wide range of scenarios.  In particular, for a given scenario, stacked survival models were found to perform better than the single model chosen through cross-validation and, at times, perform better than any single model considered in the stacking procedure.  

In practice, the true underlying data generation process is never known, i.e., one does not choose the true event distribution or functional form of the covariates.  This motivates an adaptive approach that can perform well in a wide variety of situations.  Cross-validation is currently the most common adaptive approach.  Yet, the set of simulations illustrate that stacked survival models perform as good, or better, than the model selected through cross-validation, which picks a single model to receive all the weight (i.e., $\alpha_k = 1$ for some $k$).  As such, stacked survival models warrant consideration whenever cross-validated models are used.  Other predictive models could also have been considered, though stacked survival models could inherit any particular advantages of such models through inclusion in the stack.

As shown in Section \ref{bias:var:sec}, the MSE decomposition of the stacked survival model depends on the MSE for each candidate model, the pairwise correlation between candidate models, and the weight (i.e., $\alpha_k$) given to each model.  The correlation term suggests that including an additional model that is very similar, e.g., a model of the same type as one already in the stack but with small differences in the covariates included, may not improve and could harm performance (see Supplementary Materials).  While the estimated weights can theoretically be zero, there is no guarantee that this will occur for a highly correlated model.  This could motivate a preliminary screen procedure for candidate survival models that is based on pairwise correlations.  For example, a procedure to determine the covariate combination and functional form for parametric or semi-parametric models, or determine values of tuning parameters for non-parametric estimators.  However, as noted by an anonymous referee, any screening procedure needs to be careful to avoid overfitting by using, for example, nested cross-validation or independent, external data.  The advantages of a potential screening procedure deserve further research.


Covariate dependent stacking (or, allowing the $\alpha_k$ to depend upon $\boldsymbol{x}$) is a potential avenue for improving stacked survival models.  \citet{leblanc:tibshirani:1996} mention this approach for uncensored stacking, and a collaborative group using covariate dependent stacking won the Netflix Prize competition to improve movie recommendations \citep{sill:others:2009}.  However, extending the stacking procedure to include covariate dependent weights with the constraints introduced here is not straightforward.  For example, \citet{sill:others:2009} do not constrain their covariate dependent weights despite prior experiences suggesting regularization improves performance \citep{breiman:stacking:1996, leblanc:tibshirani:1996}.  Investigation of covariate dependent stacking and different approaches to constraining the covariate dependent weights deserves further investigation.

The Brier Score, used to estimate the weighted combination of survival models, is essentially an inverse probability-of-censoring weighted (IPCW) estimate of prediction error.  The IPCW estimate requires estimating the (possibly conditional) censoring distribution.  The simulation scenarios introduced in Section \ref{sims:sec} use a Kaplan-Meier estimator for the censoring distribution that is correctly specified.  In our experience, the stacking procedure maintains good operating characteristics when the censoring model is misspecified.  However, if there is strong evidence of differential censoring among the covariates, then a conditional estimator may be warranted.

The importance of efficient, yet robust, estimators of conditional survival functions (or, equivalently, conditional distribution functions) continues to grow.  Methods in a wide range of areas require estimating a conditional survival function as a nuisance parameter, for example, censored quantile regression \citep{wey:wang:rudser:2014}, time-dependent ROC curves \citep{zheng:heagerty:2004}, inverse probability-of-censoring weighted estimators, e.g., \citet{fine:gray:1999}, model-free contrast approaches \citep{rudser:leblanc:emerson:2012}, and dynamic treatment regime methods \citep{zhao:others:2011}.  The simulations presented here suggest that stacking parametric, semi-parametric, and non-parametric models for the nuisance parameter will likely result in better estimation of regression parameters of interest, though these topics warrant further investigation.

\section*{Supplementary Materials}
The German breast cancer study data is available at:\\ {\it http://www.umass.edu/statdata/statdata/index.html}.

\section*{Acknowledgements}
This work was supported by grants UL1TR000114 of the National Center for Advancing Translational Sciences; U54MD007584 of the National Institute on Minority Health and Health Disparities; and G12MD007601 of the National Institute on Minority Health and Health Disparities.

\bibliographystyle{biorefs}
\bibliography{paperbib}

\begin{table}[hptb]
\begin{center}
\caption{Simulation results for Section \ref{sims:sec:low:cen} ($n = 200$, $p = 8$ covariates, and $25\%$ censoring) presented in integrated squared survival error (ISSE) over the observed support.  Each simulation is replicated $2000$ times, and the error is multiplied by $10$.  The two top estimators are bolded for each simulation scenario. `RSF' stands for random survival forests, `Stacking' is stacked survival models, and `CV' is the estimator selected through cross-validation.}
\begin{tabular}{lllccc}

& & & & &  \\
& & Models 	& log-Normal & Weibull & Gamma    \\ \hhline{======}

		& 	 	& log-Normal 		& {\bf 0.35} & 0.82 & {\bf 0.34}  \\
		& Single	& Weibull			& 0.61 & {\bf 0.53} & 0.41  \\
Linear	& Models	& Cox			& 0.86 & 0.68 & 0.69  \\
Effects	& 		& RSF			& 7.26 & 4.88 & 7.36  \\ \cline{2-6}

		& Flexible	& Stacking		& {\bf 0.42} & {\bf 0.58} & {\bf 0.37}  \\
		& Models	& CV			& 0.72 & 0.70 & 0.53  \\ \hhline{======}

			& 	 	& log-Normal 		& 4.71 & 2.54 & 5.03  \\
			& Single	& Weibull			& 5.17 & {\bf 2.27} & 5.30  \\
Non-Linear	& Models	& Cox			& 5.15 & 2.33 & 5.33  \\
Effects		& 		& RSF			& {\bf 4.29} & 3.49 & {\bf 4.46}  \\ \cline{2-6}

			& Flexible	& Stacking		& {\bf 3.49} & {\bf 2.08} & {\bf 3.69}    \\
			& Models	& CV			& 5.00 & 2.48 & 5.18    \\ \hhline{======}

\end{tabular}
\label{tab:sec:low:cen}
\end{center}
\end{table}

\begin{table}[hptb]
\begin{center}
\caption{Average weights for the individual models included in the stacked survival model for each of the six scenarios in Section \ref{sims:sec:low:cen} ($n = 200$, $p = 8$ covariates, and $25\%$ censoring).  Each simulation is replicated $2000$ times.  `RSF' stands for random survival forests.}
\begin{tabular}{llccc}

& & & &  \\
& Stacked Models 	& log-Normal & Weibull & Gamma    \\ \hhline{=====}

		& log-Normal 		& 0.61 & 0.19 & 0.42   \\
Linear	& Weibull			& 0.23 & 0.45 & 0.37   \\
Effects	& Cox			& 0.14 & 0.31 & 0.19   \\
		& RSF			& 0.02 & 0.05 & 0.02   \\ \hhline{=====}

			& log-Normal 		& 0.34 & 0.12 & 0.27   \\
Non-Linear	& Weibull			& 0.14 & 0.28 & 0.18   \\
Effects		& Cox			& 0.06 & 0.21 & 0.08   \\
			& RSF			& 0.46 & 0.39 & 0.47   \\ \hhline{=====}

\end{tabular}
\label{tab:low:cen:ave:wghts}
\end{center}
\end{table}

\begin{table}[hptb]
\begin{center}
\caption{Simulation results for Section \ref{sims:sec:high:cov} ($n = 200$, $p = 80$ covariates, and $25\%$ censoring) presented in integrated squared survival error (ISSE) over the observed support.  Each simulation is replicated $2000$ times, and the error is multiplied by $1$.  The two top estimators are bolded for each simulation scenario. `RSF' stands for random survival forests, `Stacking' is stacked survival models, and `CV' is the estimator selected through cross-validation.}
\begin{tabular}{lllccc}

& & & & &  \\
& & Models 	& log-Normal & Weibull & Gamma    \\ \hhline{======}

		& Single	& Cox - Lasso		& {\bf 2.43} & {\bf 1.68} & {\bf 2.50}  \\
Linear	& Models	& Cox - Boosting	& 2.60 & 1.75 & 2.66  \\
Effects	& 		& RSF			& 2.46 & 1.86 & {\bf 2.50}  \\ \cline{2-6}

		& Flexible	& Stacking		& {\bf 2.43} & {\bf 1.68} & {\bf 2.50}   \\
		& Models	& CV			& {\bf 2.43} & 1.69 & {\bf 2.50}   \\ \hhline{======}

			& Single	& Cox - Lasso		& 2.02 & 1.03 & 2.08   \\
Non-Linear	& Models	& Cox - Boosting	& 2.01 & {\bf 1.01} & 2.08   \\
Effects		& 		& RSF			& {\bf 1.89} & 1.15 & {\bf 1.95}   \\ \cline{2-6}

			& Flexible	& Stacking		& {\bf 1.97} & {\bf 1.00} & {\bf 2.04}   \\
			& Models	& CV			& 2.01 & 1.03 & 2.07   \\ \hhline{======}

\end{tabular}
\label{tab:sec:high:cov}
\end{center}
\end{table}

\end{document}